\documentclass[preprint2]{aastex}

\newcommand{\heo}{\ion{He}{2} $\lambda$1640 + \ion{O}{3}] $\lambda$1663}
\newcommand{\car}{\ion{C}{4} $\lambda$1549}
\newcommand{\lya}{Ly$\alpha$}
\newcommand{\sio}{\ion{Si}{4} $\lambda$1400 + \ion{O}{4}] $\lambda$1402}
\newcommand{\sic}{\ion{Si}{3}] $\lambda$1892 + \ion{C}{3}] $\lambda$1909}
\newcommand{\hb}{H$\beta$}

\shortauthors{Onken \& Peterson}
\shorttitle{A Supermassive Black Hole in NGC 3783}

\received{}
\accepted{}

\begin{document}

\title{The Mass of the Central Black Hole in the Seyfert Galaxy NGC 3783}

\author{Christopher A. Onken \& Bradley M. Peterson }

\affil{Department of Astronomy, The Ohio State University, Columbus, OH 43210}
\email{onken, peterson@astronomy.ohio-state.edu}

\begin{abstract}

Improved analysis of ultraviolet and optical monitoring data on the Seyfert 1
galaxy NGC 3783 provides evidence for the existence of a supermassive, 
(8.7$\pm$1.1)$\times10^{6}$ M$_{\odot}$, black hole in this galaxy.
By using recalibrated spectra from the 
{\it International Ultraviolet Explorer} satellite and ground-based optical data,
as well as refined techniques
of reverberation mapping analysis, we have reduced the statistical uncertainties 
in the response of the emission lines to variations in the ionizing continuum.
The different time lags in the emission line responses indicate a stratification in the  
ionization structure of the broad-line region and are
consistent with the virial relationship suggested by the analysis of similar active
galaxies.

\end{abstract}


\clearpage

\section{INTRODUCTION}

The primary model that has emerged over the last few decades for the radiation source 
of an active galactic nucleus (AGN) is accretion onto a supermassive 
black hole (SMBH). Variations in the ionizing continuum have been seen to influence the
strength of emission lines arising from the broad-line region (BLR).
Cross-correlation of the continuum and emission line light curves yields a 
characteristic time lag with which each line echoes the continuum fluctuations
\citep{bla82}.
This reverberation mapping technique has been used to measure the sizes of BLRs
for a growing number of AGNs (see Wandel, Peterson, \& Malkan 1999; Kaspi et al. 2000).

In addition to the BLR 
size, reverberation analysis can be used to estimate the mass of the SMBH. The
reliability of these reverberation masses has been debated because of the uncertainty
surrounding the common assumption of virialized BLR gas motions. The 
detailed kinematics and structure of the BLR is an unresolved issue and could
lead to systematic errors on the order of a factor of a few or perhaps more
\cite[see][]{fro01,kro01}. However, the excellent agreement between the black hole
mass-bulge velocity dispersion (M-$\sigma$) relationships for reverberation-mapped AGNs and
normal galaxies \citep{fer01} suggests the systematic discrepancy 
introduced in reverberation mapping is small. Additionally, AGNs for which multiple
emission lines have been mapped (NGC 5548, 3C 390.3, NGC 7469) 
show an inverse relationship between the time lag
and the emission line width, consistent with the gas motions being dominated by the gravity
of the SMBH \citep{pet99,pet00}.

A combined optical and UV monitoring campaign was carried out on the Seyfert 1 galaxy
NGC 3783 by the {\it International AGN Watch} consortium, making use of the 
{\it International Ultraviolet Explorer} ({\it IUE}), the {\it Hubble Space
Telescope}, and a host of ground-based observatories over a period of 7 months 
in 1991-1992. The results of that work have been published by \citet{rei94}, \citet{sti94},
and \citet{all95}. Compared to the consortium's earlier study of another Seyfert galaxy, NGC 5548
\cite[see][]{cla91,pet91,die93}, the emission-line time lags were relatively uncertain,
too poorly constrained in fact to reveal any possible virial relationship between line
width and time lag.

The continued rarity of such campaigns, however, makes it clearly desirable to learn
as much as possible from the extant datasets. This provides the motivation for our
current study.

With the release of an updated processing
pipeline and calibration for {\it IUE} data, the possibility arose to re-analyze the spectra
of NGC 3783 and reduce the uncertainties of the emission-line time lags. In addition, the
techniques of reverberation analysis have matured in the years since the original
data were published, now providing more consistent methodology for cross-correlation 
and error estimation. Thus, we have re-examined the data, deriving more precise
results for the emission line reverberation and revising the previous estimates of
the reverberation mass.

The next section describes the observations and how the data were reduced (\S2). In \S3 we
explain the analysis procedure and give our cross-correlation results. Section 4 discusses the
results and the SMBH mass determination, and our conclusions are summarized in \S5.

\section{OBSERVATIONS AND DATA REDUCTION}

\subsection{UV Data} \label{UVobs}

The {\it IUE} observations of NGC 3783 were conducted in 69 separate epochs, with two
sampling rates. The first interval (of 45 epochs) had an average spacing of 4.0 days,
while the final 24 epochs observed the AGN with an average spacing of 2.0 days. A more
complete description of the UV observing program is provided by \citet{rei94}.

In addition to the original {\it IUE} Spectral Image Processing System (IUESIPS), 
\citet{rei94} used a Gaussian extraction method \cite[GEX; see][]{cla91} to obtain
the spectra of NGC 3783.
After the original data had been taken, a new standard processing pipeline was introduced. 
The main advantages of the New Spectral Image Processing System \cite[NEWSIPS;][]{nic93}  
with respect to the older IUESIPS are 
the improved photometric accuracy and higher $S/N$ of 
the spectra; 
these characteristics have been achieved by introducing a new method 
of raw data science registration (which both reduces the fixed pattern 
noise in the images and improves the photometric corrections), a 
weighted slit extraction method, and re-derived absolute flux calibrations. 
NEWSIPS also includes corrections for non-linearity that might have 
affected previous studies. Overall, NEWSIPS-processed spectra show average $S/N$
increases of 10--50\% over IUESIPS data \citep{nic96}.

We retrieved the NEWSIPS-extracted short wavelength prime camera 
\citep[SWP;][]{har87} spectra from the {\it IUE Final 
Archive}\footnote{\url{http://ines.laeff.esa.es/ines/}}.
While \citet{rei94} analyzed data from both the SWP and long wavelength prime cameras,
we have limited our study to observations made with the SWP
instrument, which has a wavelength range of 1150--1975 \AA\ in the low-dispersion mode 
\citep{new92}. 

Each spectrum was examined and several types of problems led to spectra
being removed from further consideration: (1) low $S/N$ (determined by inspection, but corresponding
roughly to a continuum $S/N$ limit of 10); (2) 
unusual spectral features (possibly due to grazing cosmic-ray impacts); (3) short exposure times
(when longer-exposure data were available from the same epoch and the line flux data were discrepant).
Some anomalous features were checked against the GEX frames, from which cosmic ray impacts
were carefully removed.
Problems with the spectra were ignored in cases where they occurred in spectral regions outside
those used in computing line and continuum fluxes. 
Continuum and emission line flux values were measured using the 
wavelengths limits listed in Table \ref{tab1}. 

The continuum was defined by a linear fit through four spectral regions (1340--1370 \AA, 
1440--1480 \AA, 1710--1730 \AA, and 1840--1860 \AA). An alternate fit through the first
three of these regions produced consistent results. Wavelength-specific problems in two
cases (SWP 45150, SWP 45206) led us to substitute the alternate continuum fit for these
spectra.

We have estimated the flux uncertainties by considering instances in which multiple
independent exposures were obtained at the same epoch (i.e., a single pointing toward the 
target). Flux ratios between pairs of points within each epoch were calculated and the
standard deviation of the flux ratios was taken as the fractional uncertainty for all
observations.
This analysis was conducted independently for each emission line and continuum band.
As noted above, however, highly discrepant data were
removed prior to this analysis. In spite of our use of an edited dataset, 
the large number of data pairs contributing
to our error estimate (about 35) justifies our continued use of these values in the analysis.
The final UV dataset is given in Table \ref{tab2} for the continuum measurements and
in Table \ref{tab3} for the emission lines.

The velocity width desired for the reverberation mass calculation is related to the 
emission line velocity full width at half-maximum ($V_{FWHM}$) by
\begin{equation} \label{equ1}
\sigma = \frac{\sqrt{3} V_{FWHM}}{2},
\end{equation}
where the factor of $\sqrt{3}/2$ is used to maintain consistency with previous work
\cite[e.g.,][]{wan99,kas00} and assumes isotropic gas motion.

An RMS spectrum was created from the data to isolate
the varying parts of the emission lines
and it was from this spectrum that the primary $V_{FWHM}$ values for the emission
lines were measured. 
The $V_{FWHM}$ data were constructed by considering the extreme flux values within
the continuum regions, fitting two continuum slopes (to the highest flux levels and lowest flux
levels), and averaging the measures of $V_{FWHM}$ derived from the two continuum determinations. 
Finally, the data were converted to their rest-frame widths using $z$ = 0.009730 $\pm$ 0.000007
\citep{the98}. 
Previous work examining the difference between using the mean and
RMS spectra have not produced significantly different results \cite[e.g.,][]{kas00}, but in
principle the RMS spectrum should better trace the gas with which we are concerned. 
We have measured line widths from both spectra (Figure \ref{fig1}) and report the
results of our mean and RMS $V_{FWHM}^{rest}$ measurements in Table \ref{tab4}. 
Geocoronal \lya\ emission blended into
the \lya\ spectral region precludes $V_{FWHM}$ measurement for this line and thus also prevents \lya\
contribution to the mass determination, but cross-correlation analysis is still feasible
by excluding the contaminated portion of the spectrum (see Table \ref{tab1}).

\subsection{Optical Data}

Ground-based optical spectroscopy was conducted over the same time period as the {\it IUE} observations.
The optical data analyzed here were retrieved from the {\it AGN Watch} 
website\footnote{\url{http://www.astronomy.ohio-state.edu/\~{}agnwatch/}}, and details of
the observations are described by \citet{sti94}. 
We have limited our investigation to the data
gathered at the Cerro Tololo Inter-American Observatory (CTIO) 1.0 m telescope 
to ensure the most homogeneous dataset 
possible for the cross-correlation analysis and for the construction of the mean and
RMS spectra.
Spectra that were excessively noisy or contained other anomalies were discarded from 
consideration, leaving 37 CTIO observations for further analysis.

Narrow spectral lines are assumed not to vary over the timescales these data are 
probing. Thus the individual spectra were scaled to a constant flux by using
the spectral scaling technique of
\citet{van92}. This method computes a smooth scaling function between the input spectrum and a
reference (the mean spectrum in this case, shown in Figure \ref{fig2}) 
over a specified wavelength range. We scaled over the spectral region
4972--5150 \AA\ in order to span the redshifted [\ion{O}{3}] $\lambda\lambda$4959, 5007 emission lines
and a suitable amount of continuum. 
We found that two iterations were required for full convergence.
This reduced the fractional RMS scatter in the [\ion{O}{3}] $\lambda$5007 light curve
(measured between 5028 and 5090 \AA) to less than 2.5\%. 
Additional iterations failed to produce any light curves with smaller scatter. 
The mean [\ion{O}{3}] $\lambda$5007 flux was 
normalized to 8.44$\times$10$^{-13}$ erg s$^{-1}$ cm$^{-2}$, the value derived by the careful
analysis of \citet{sti94}.

Following the calibration of the spectra, the \hb\ line was measured between 4830 and 
4985 \AA\ (with the continuum set by a linear fit between 4800--4820 \AA\ and 5130--5150 \AA). The 
flux uncertainties
were measured in the same way as for the UV data (\S\ref{UVobs}), and the results are
given in Table \ref{tab5}. Due to the smaller optical dataset,
the flux errors for \hb\ and the 5150 \AA\ continuum rely on only seven data pairs. 
To be cautious, we have been more conservative in our estimation of the optical flux uncertainties.
The method for measuring $V_{FWHM}^{rest}$ was also applied to the optical data and yielded values of
(2.91$\pm$0.19)$\times$10$^{3}$ km s$^{-1}$ for the RMS spectrum and (2.65$\pm$0.02)$\times$10$^{3}$
km s$^{-1}$ for the mean optical spectrum.

\section{LIGHT CURVE ANALYSIS}

In Table \ref{tab6} we compare the sampling 
characteristics of our data with the previously published light curves. 
When we bin the data in each epoch, the variability parameters of the old and
new UV datasets appear nearly identical. The ``excess variance'', $F_{var}$, represents the
mean fractional variation of each dataset \cite[see][]{rod97}; $R_{max}$ is the ratio of maximum to
minimum flux levels. The updated optical dataset is much more sparse than the
previously published data because of our desire for the most homogeneous dataset
possible.

Figure \ref{fig3} shows the light curves for each of the UV and optical emission lines
and continuum bands.
Applying the techniques described by \citet{pet98}, 
we generated cross-correlation functions (CCFs) relating the various 
emission line light curves to the 1355 \AA\ continuum flux. 
We report both peak ($\tau_{peak}$) and centroid ($\tau_{cent}$) cross-correlation lags.
However, the reader should be warned that ``lags'' in the text will hereafter refer 
to centroids, unless otherwise noted, and that such lags do not
represent a simple phase shift between the light curves.

As \citet{kor91a} noted for NGC 3783 (and other reverberation-mapped AGNs), 
the choice of what threshold to use for the
centroid calculation can significantly affect the resulting time lag.
Figure \ref{fig4} shows that some lines tend toward larger lags and others toward smaller
values as
the centroid becomes increasingly dominated by the peak value. 
For the interpolated CCF \cite[ICCF;][]{gas87, whi94}, 
we experimented
with different interpolation lengths and different thresholds for the calculation of the
lag centroid. Our subsequent analysis uses an interpolation unit of 0.1 days in both 
light curves (interpolating one dataset at a time, with the resulting lags averaged) 
and a centroid threshold of 80\% of the peak correlation coefficient. 

In addition to the ICCF, we calculated the discrete correlation function \cite[DCF;][]{ede88}
for each continuum band and emission line. While the DCF, which requires binning of the data,
is more likely to miss a real correlation
than the ICCF under poor sampling conditions, it is also less likely to introduce a spurious 
relationship \citep{whi94}. \citet{gas94} notes that the DCF also relies on interpolation, but does so in 
the correlation function, rather than the original time series.
The ICCF and DCF methods have been compared by various authors 
\cite[e.g.,][]{whi94, lit95} and typically yield similar results. 

To assess the uncertainties in the time lag calculations, we used the Monte Carlo (MC) methods
of \citet{pet98}. This technique for model-independent error
estimation consists of two components, each testing for a separate contribution to the
cross-correlation uncertainty. 
To account for the uncertainty in an individual flux measurement, each data point in the light
curve is altered
by a random Gaussian deviation that corresponds to the quoted flux error (calculated by the
method described in \S\ref{UVobs}). The
result of many such realizations, referred to as ``flux randomization'' (FR), should 
yield average values equal to the original data
with standard deviations given by the original uncertainties. 
Secondly, the effects of non-uniform temporal sampling of the AGN fluctuations are 
investigated with ``random subset selection'' (RSS). Given a sample
of $N$ observations, $N$ data points are randomly chosen from the set (ignoring whether 
they have been chosen previously). While DCF and ZDCF \citep{ale97} analyses can weight multiply-selected
data, the ICCF (which we use for our MC calculations) does not consider the 
flux uncertainties and simply excludes the
redundant data points. Ignoring these data reduces the set by $\sim$$N/e$ on average and
so should yield a wider range of peak lags from the ICCF.
Repeated MC realizations (at least 10$^{3}$ in the present work; combining the FR/RSS
methods for each calculation) are used to create
a cross-correlation peak distribution \cite[CCPD;][]{mao89}, which provides an empirical
measurement of the uncertainties for both $\tau_{cent}$ and $\tau_{peak}$.

\subsection{Emission Lines}

The light curves for each emission line (\heo, \sio, \lya, \car, \sic, and \hb)
were run through the ICCF, DCF, and FR/RSS programs, using the 1355 \AA\ continuum
data as the ``driving'' light curve.
The CCFs and CCPDs are shown for each emission line
in the panels of Figure \ref{fig5}. The CCPDs are shown to give a graphical 
indication of the empirical uncertainties and are scaled to the maximum value in
each panel. 

Each of the emission lines was very well correlated with the continuum flux.
The poorest correlation with the continuum was found for \sic, which was found to have a 
peak ICCF value of r$_{max}$=0.354 (i.e., a probability of arising from an uncorrelated parent 
population of roughly $<$ 0.001) and we limited the range of computation for this line to $\pm$16 days to
avoid aliasing. Table \ref{tab7} summarizes the previous data and our new results. 

Our results for the UV emission lines are generally in agreement with those of \citet{rei94}.
It should be noted, however, that the peak and centroid lags calculated by \citet{rei94} and
\citet{sti94} used the 1460 \AA\ continuum as the driving light curve. The results quoted
here are consistent with those derived from the recalibrated data with the continuum 
centered at 1460 \AA\ rather than at 1355 \AA. 
Because of the large uncertainties assigned to previous lag values, most of our NEWSIPS lags 
are within 1-$\sigma$ of the old data. The exceptions are \sic, for which the IUESIPS-based data
failed to produce any lag at all, and \hb. The GEX extraction method yielded a peak lag for
\sic\ similar to what we found, but a centroid lag approximately 2-$\sigma$ larger than 
the current result. Our centroid lag for \hb\ was only slightly more than 1-$\sigma$ 
greater than the previous value.

The significant discrepancy between our \hb\ results and those of 
\citet[][4.5$^{+3.6}_{-3.1}$ days]{wan99} arises
from the double-peaked nature of the CCF. The centroid lag 
calculated for the 5150 \AA\ continuum-\hb\ CCF is based on fewer points than the
1355 \AA\ continuum-\hb\ lag, and gives precedence to the peak at smaller lags. We
have greater confidence in the results that use the UV continuum data, and those results closely
match the UV-\hb\ correlation found by \citet{sti94}.

\subsection{Continuum}

Strong evidence for wavelength-dependent continuum lags has been found for
only two AGNs (NGC 7469 and Akn 564), but appears to be 
consistent with simple accretion disk models that predict $\tau \propto \lambda^{4/3}$
\cite[see][]{wan97,col98,col01}. However, \citet{kor01} note that diffuse emission 
from broad-line clouds can produce a similar wavelength dependence, so the
origin of this phenomenon is not clear.

The large uncertainties still present in the NEWSIPS continuum lags prevent us from reasonably
testing the $\tau$--$\lambda$ relationship because the continuum-continuum time
lags we find are not statistically significant (see Table \ref{tab7}).

\section{IMPLICATIONS FOR THE BLR AND THE SMBH}

As the tabular data indicate, the expected pattern of more highly ionized lines 
having smaller time lags (i.e., originating closer to the ionization source)
is reconfirmed by our analysis. 

Figure \ref{fig6} plots $V^{rest}_{FWHM}(RMS)$ versus $\tau^{rest}_{cent}$ 
for the five emission lines we measured.
The virial assumption predicts a slope of $-0.5$ (in log-log space). Deviation from this
relationship would contradict our model, but agreement with the predicted slope
cannot rule out other dynamical possibilites \cite[see][and references therein]{kro01}.

The statistical problem of fitting to intrinsically scattered data with heteroscedastic errors
has been addressed with computational methods by \citet{akr96}. However, our data has the
additional difficulty of asymmetric errors in the lags. 
To account for the asymmetric time lag uncertainties we first used the larger of the two lag
errors and then assessed in which direction the data points differed from the regression.
We recalculated the fit using the errors toward the previous regression and confirmed that those were
the appropriate choices in the final fit. The slope of the $V^{rest}_{FWHM}(RMS)$-$\tau^{rest}_{cent}$
relation derived by the regression software\footnote{available at
\url{http://www.astro.wisc.edu/\~{}mab/archive/stats/stats.html}} was $-0.450\pm0.070$, consistent
with our expectations for a virial relationship (irrespective of the specific multiplicative 
factor relating the line widths and $V_{FWHM}$ values). Hence, we
fixed the slope at $-0.5$ and calculated the mass independently for each emission line, applying 
our previously stated assumption of isotropic BLR gas motion and inserting the
appropriate rest-frame values into the following equation:
\begin{equation}
M = \frac{3\ c\ \tau\ V_{FWHM}^{2}}{4\ G}.
\end{equation}
Weighting the data by the uncertainty in the direction of the mean (since the lag errors are
still asymmetric) yields an average SMBH mass of (8.7$\pm$1.1)$\times10^{6}$ M$_{\odot}$.

Previous work with {\it IUE} archival data having much poorer temporal resolution measured
a much larger \car\ time lag and derived a mass of 7.3$^{+3.5}_{-3.6}\times10^{7}$ M$_{\odot}$
\citep{kor91a,kor91b}.
\citet{wan99} calculated the \hb\ lag with respect to the 5100 \AA\ continuum from the light
curves of \citet{sti94} and then used the RMS velocity width to estimate a mass of 
1.1$^{+1.1}_{-1.0}\times10^{7}$ M$_{\odot}$.
Applying this method to our version of the optical data yields an SMBH mass of 
6.2$^{+4.7}_{-6.1}\times10^{6} M_{\odot}$, within the 1-$\sigma$ error bars 
for our mass measurement with the full dataset. 
\citet{fro00} employed a different means of measuring the velocity dispersion from
the data of \citet{rei94} and
derived masses of 1.6$^{+0.8}_{-0.4}\times10^{7}$ M$_{\odot}$ and 
1.3$^{+0.8}_{-0.5}\times10^{7}$ M$_{\odot}$ from \lya\ and \car, respectively.
Various disk accretion models predicting a
SMBH mass in the range of 2.0--7.0$\times10^{7}$ M$_{\odot}$ were cited by \citet{all95}.
However, they note the simple nature of these spatially thin, optically thick disk models
and the potential for a large discrepancy from the true SMBH mass. 

\section{SUMMARY}

We have conducted reverberation mapping analysis on recalibrated {\it IUE} and ground-based
optical observations of
the Seyfert 1 galaxy NGC 3783 with the goal of revising the mass estimate for the central
SMBH. The NEWSIPS spectra confirm the existence of varying time lags for emission lines
of different ionization potentials and provide a better constraint on the SMBH mass under
the assumption of virial gas motion. The emission line time lags vary from 1.3 to 10.4 days, 
and analysis of peak and centroid time
lags yield similar results for each line. Our mass determination revises the previous values
to a mass of (8.7$\pm$1.1)$\times10^{6}$ M$_{\odot}$.

\acknowledgments

We gratefully acknowledge support for this work through NASA grant NAG5-8397.
C. A. O. thanks The Ohio State University for support through the Distinguished
University Fellowship. We also thank Patrizia Romano for acquiring the NEWSIPS data, and Matthew A.
Bershady for making available his statistical software.
This research has made use of the NASA/IPAC Extragalactic Database (NED) 
which is operated by the Jet Propulsion Laboratory, California
Institute of Technology, under contract with the National Aeronautics 
and Space Administration.

%figures

\clearpage

\plotone{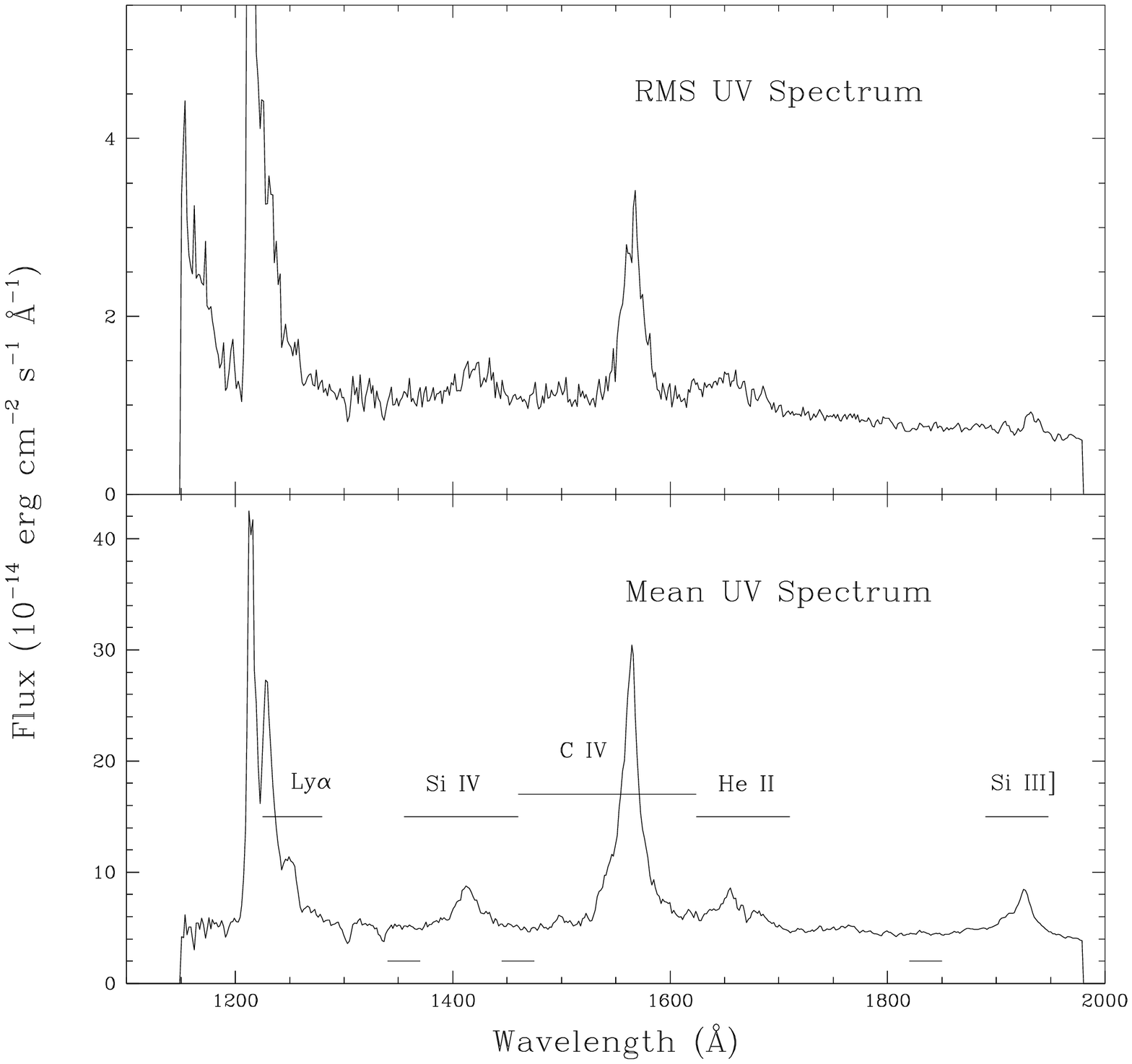}
\figcaption[f1.eps]{{\it Top}: RMS UV spectrum. 
{\it Bottom}: Mean UV spectrum. Wavelengths delineated above the spectrum
indicate emission-line ranges; those below the spectrum mark ranges of
continuum flux measurement. \label{fig1}}

\plotone{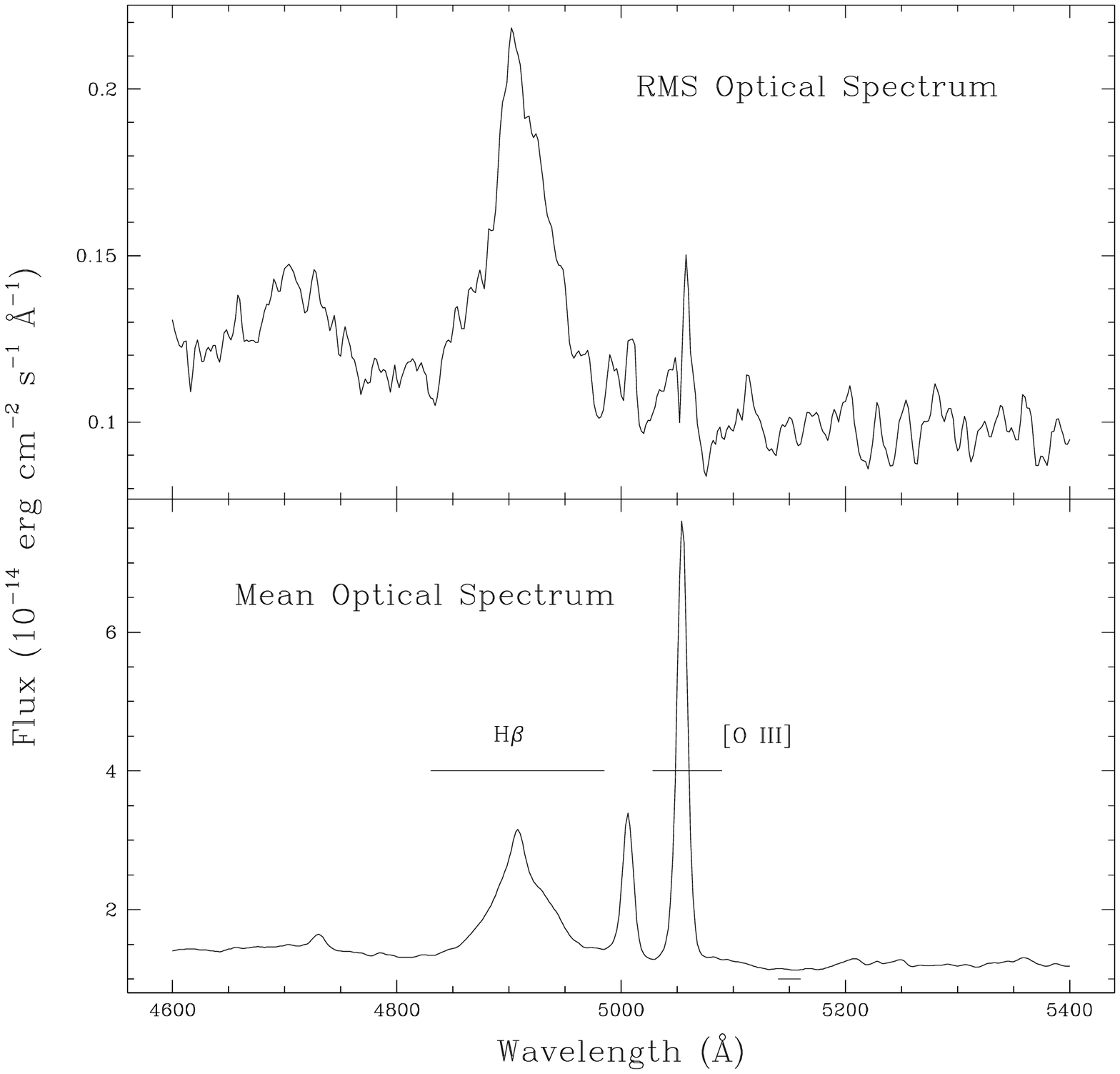}
\figcaption[f2.eps]{{\it Top}: RMS optical spectrum. {\it Bottom}: Mean optical 
spectrum. Wavelength indications as in Fig. 1. \label{fig2}}

\epsscale{0.8}
\plotone{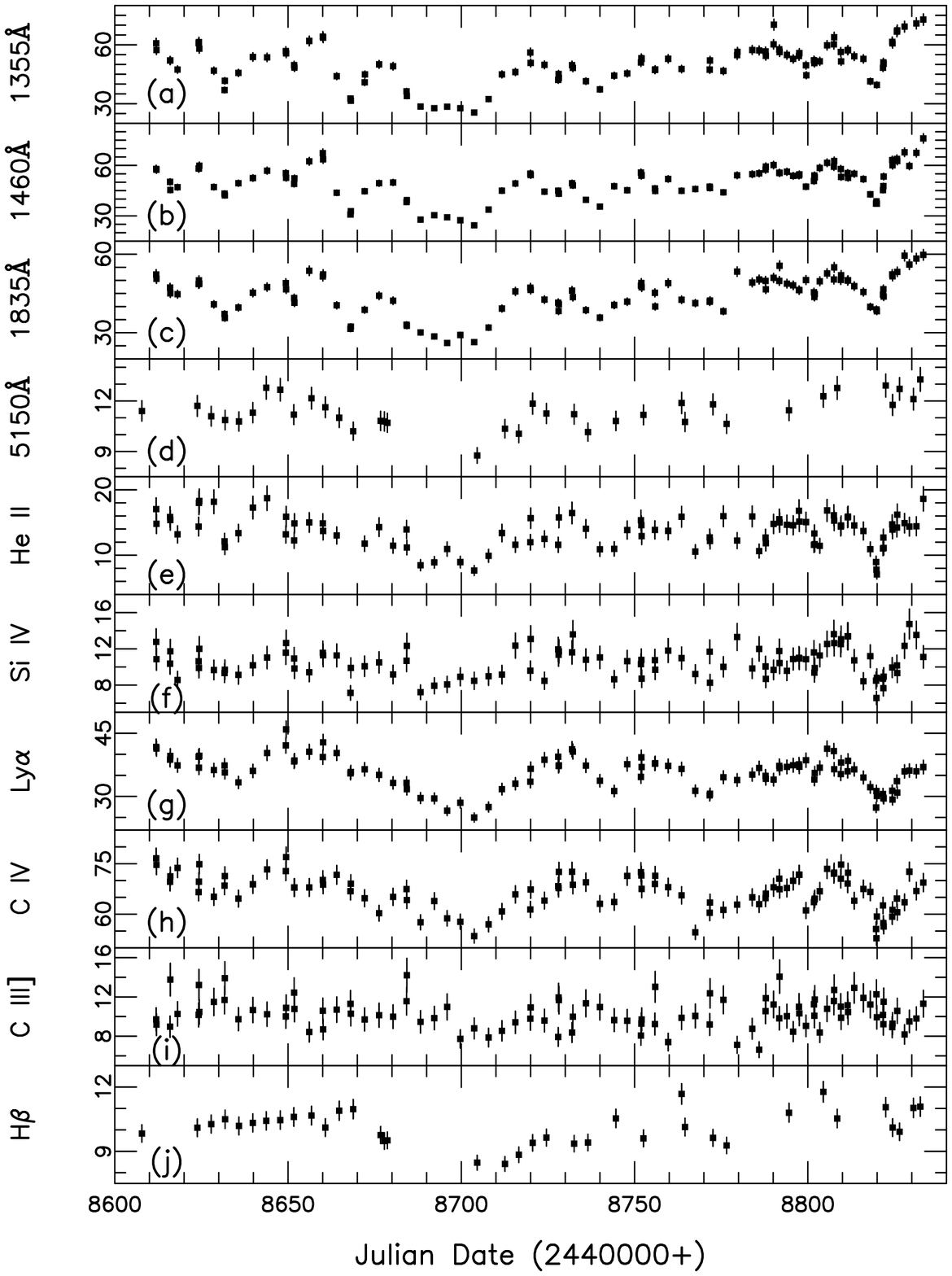}
\figcaption[f3.eps]{UV and optical light curves:
(a) 1355 \AA\ continuum,
(b) 1460 \AA\ continuum, (c) 1835 \AA\ continuum, (d) 5150 \AA\ continuum,
(e) \heo, (f) \sio, (g) \lya, (h) \car, (i) \sic, and (j) \hb. Continuum fluxes
are in units of 10$^{-15}$ ergs cm$^{-2}$ s$^{-1}$ \AA$^{-1}$. Emission line fluxes
are given in units of 10$^{-13}$ ergs cm$^{-2}$ s$^{-1}$. \label{fig3}}

\clearpage

\epsscale{0.8}
\plotone{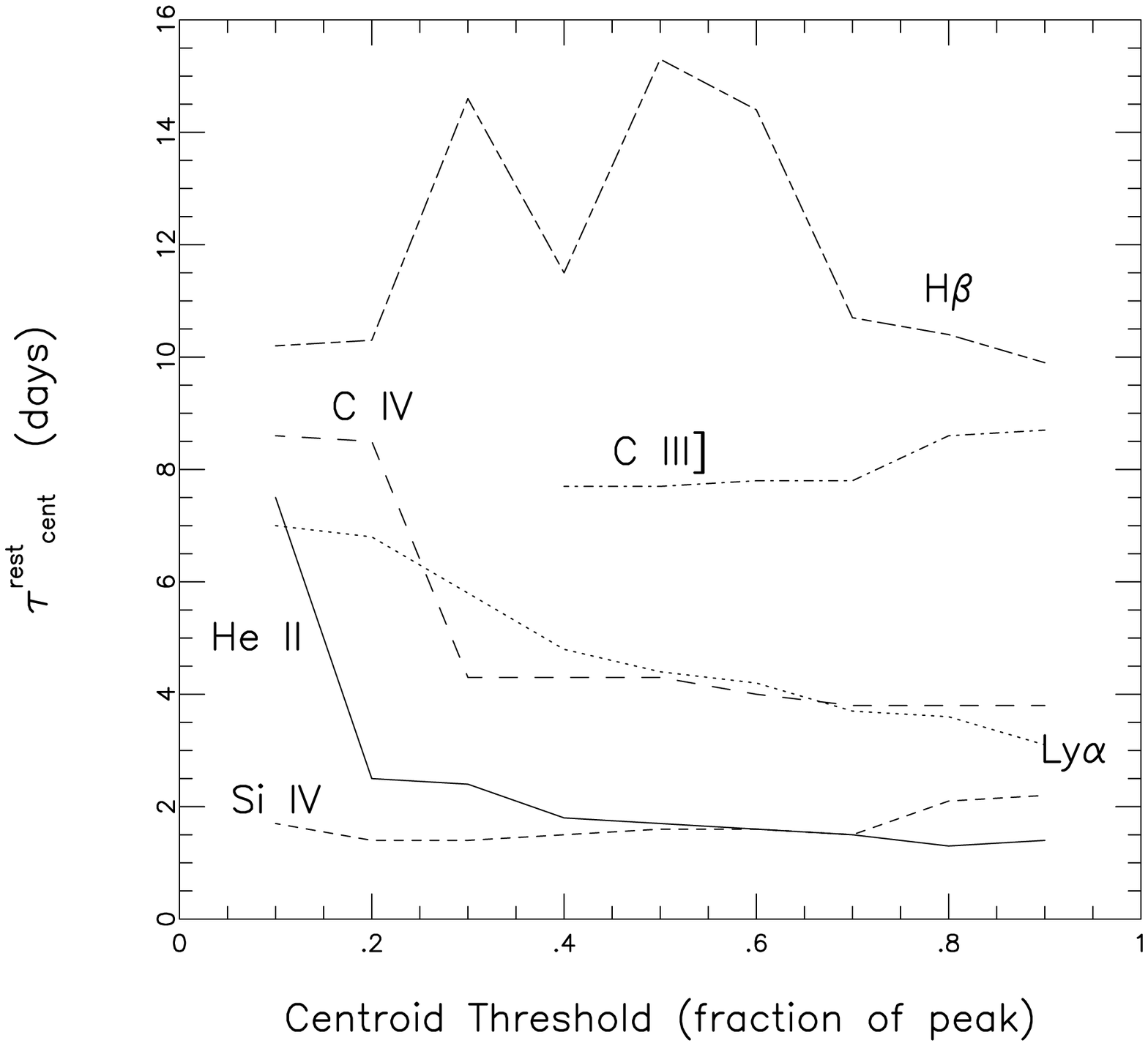}
\figcaption[f4.eps]{Rest-frame centroid time lag versus threshold level for 
centroid determination (as a fraction of the peak correlation coefficient).
The data plotted are for \heo\ (solid), \sio\ (short dashed), \lya\ (dotted),
\car\ (long dashed), \sic\ (dot-short dashed), and \hb\ (dot-long dashed). 
\label{fig4}}

\epsscale{0.75}
\plotone{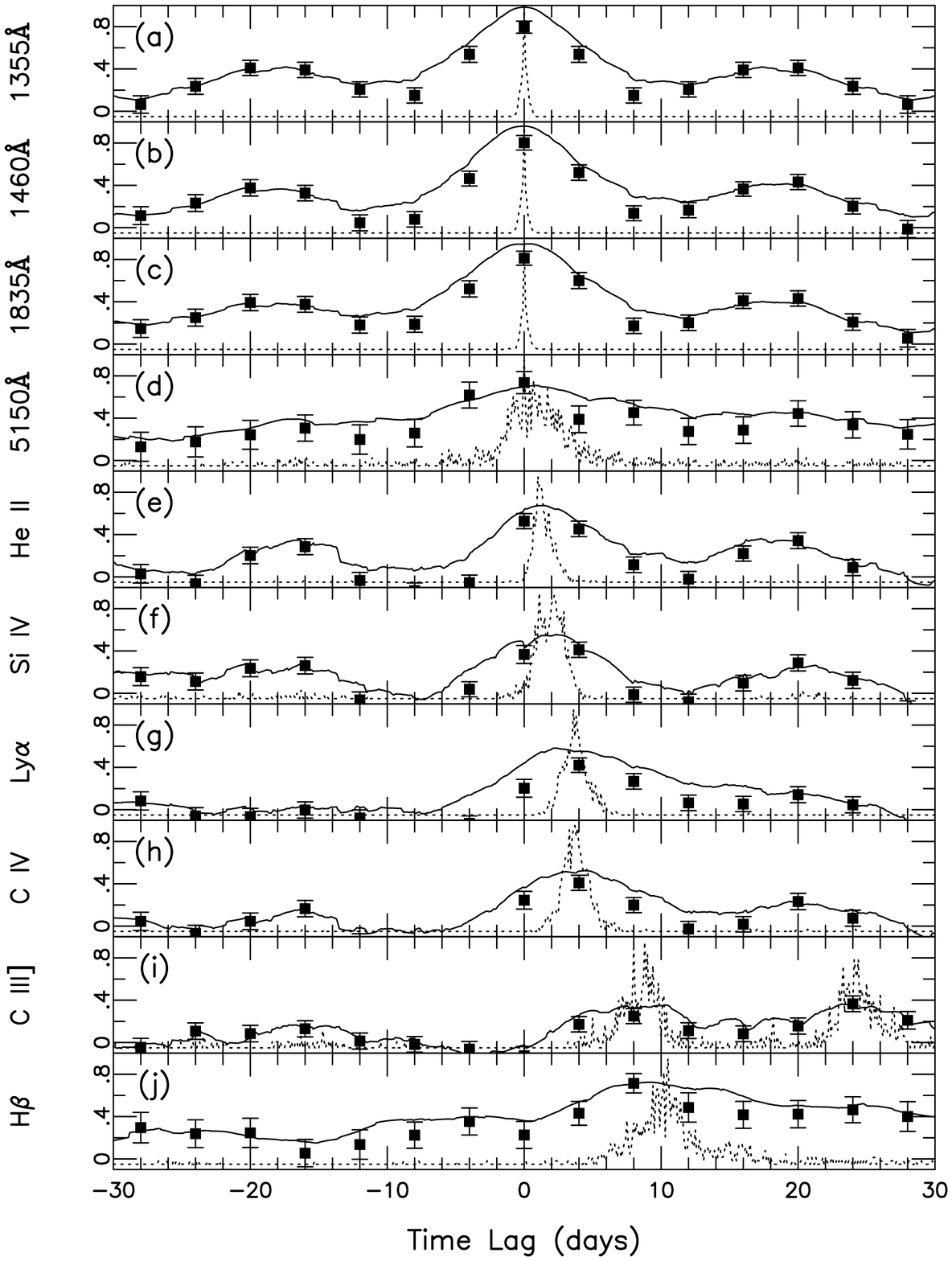}
\figcaption[f5.eps]{Results of cross-correlation of the 1355 \AA\ continuum with
(a) itself;
(b) 1460 \AA\ continuum; (c) 1835 \AA\ continuum; (d) 5150 \AA\ continuum;
(e) \heo; (f) \sio; (g) \lya; (h) \car; (i) \sic; (j) \hb.
The solid lines show the ICCFs, the data points are the DCFs, 
and the dashed lines represent the CCPDs. Note that the y-axis scale for
the CCPDs is the fraction of MC realizations producing a centroid of that
lag value and is scaled to the maximum in each panel. \label{fig5}}

\epsscale{0.8}
\plotone{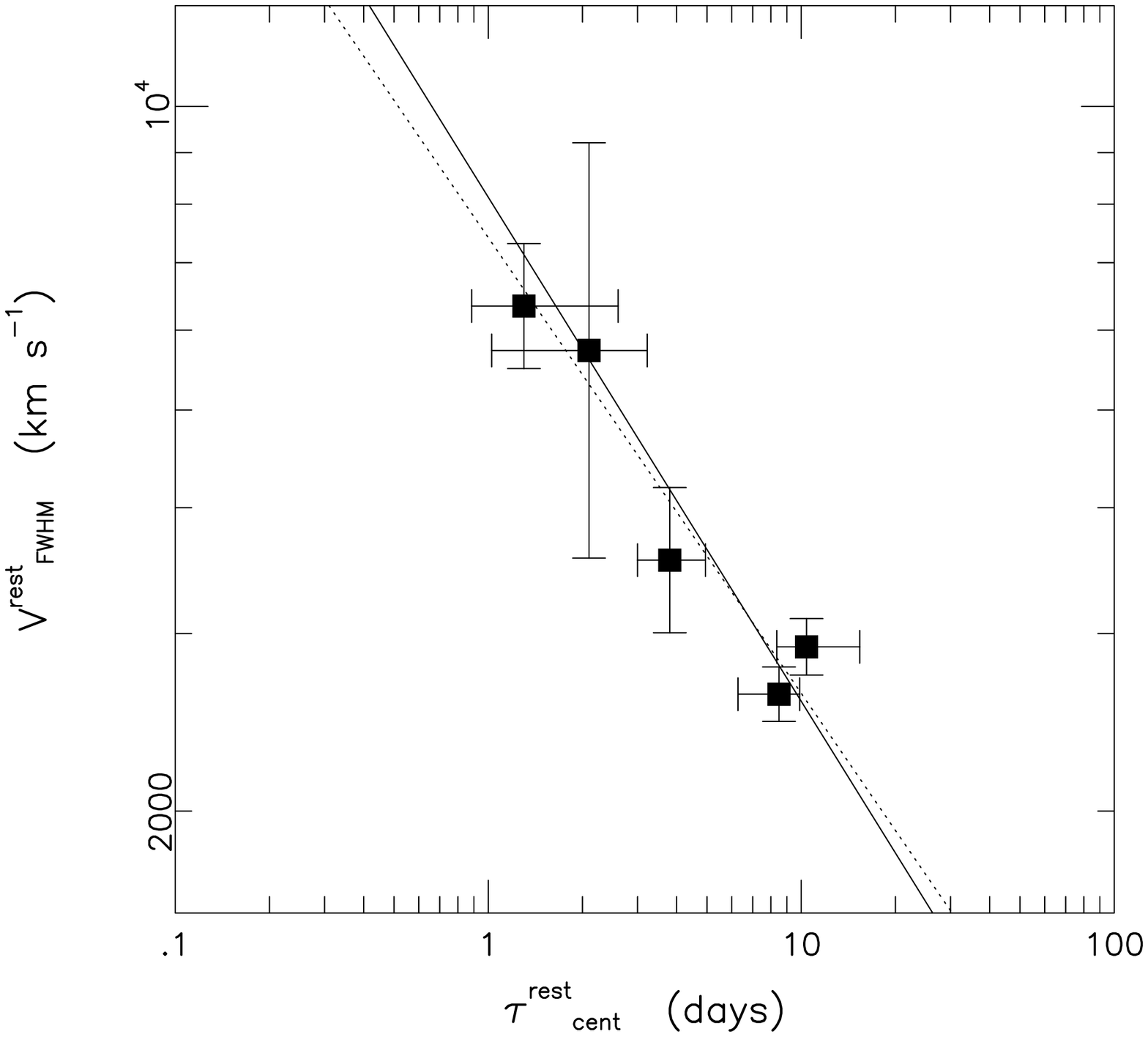}
\figcaption[f6.eps]{Rest-frame velocity FWHM versus rest-frame centroid time lag 
for the five emission lines we have measured. The dashed line is the best fit to the
data; the solid line is the best fit with fixed slope of -0.5. \label{fig6}}

\clearpage

\begin{deluxetable}{lcc}
\tablecaption{Wavelength Limits \label{tab1}}
\tablewidth{0pt}
\tablehead{
\colhead{Line/Band} & \colhead{Wavelength Range (\AA)}
}
\startdata
1355 \AA\ continuum &1340--1370 \\
1460 \AA\ continuum &1445--1475 \\
1835 \AA\ continuum &1820--1850 \\
5150 \AA\ continuum &5140--5160 \\
\lya &1225--1280 \\
\sio &1355--1460 \\
\car &1460--1624 \\
\heo &1624--1710 \\
\sic &1890--1948 \\
\hb &4830--4985 \\
\enddata
\end{deluxetable}

\clearpage

\begin{deluxetable}{lcccc}
\tablecaption{UV Continuum Flux Data\tablenotemark{a} \label{tab2}}
\tablewidth{0pt}
\tablehead{
\colhead{Image} & \colhead{Julian Date} & \colhead{} & \colhead{} & \colhead{} \\
\colhead{Name} & \colhead{(2,440,000+)} & \colhead{F($\lambda1355$)} & 
\colhead{F($\lambda1460$)} & \colhead{F($\lambda1835$)} 
}
\startdata
SWP 43438 &8611.948 &60.811$\pm$2.432 &57.899$\pm$2.142 &52.244$\pm$1.776 \\
SWP 43439 &8612.031 &57.353$\pm$2.294 &57.395$\pm$2.124 &50.582$\pm$1.720 \\
SWP 43472 &8615.949 &51.984$\pm$2.079 &50.204$\pm$1.858 &47.259$\pm$1.607 \\
SWP 43473 &8616.029 &51.979$\pm$2.079 &45.407$\pm$1.680 &45.024$\pm$1.531 \\
SWP 43485 &8618.118 &47.391$\pm$1.896 &47.043$\pm$1.741 &44.713$\pm$1.520 \\
SWP 43539 &8624.202 &60.846$\pm$2.434 &58.123$\pm$2.151 &48.605$\pm$1.653 \\
SWP 43540 &8624.294 &61.371$\pm$2.455 &58.173$\pm$2.152 &50.031$\pm$1.701 \\
SWP 43541 &8624.385 &58.033$\pm$2.321 &59.779$\pm$2.212 &48.735$\pm$1.657 \\
SWP 43557 &8628.595 &46.814$\pm$1.873 &47.146$\pm$1.744 &40.850$\pm$1.389 \\
SWP 43587 &8631.680 &36.975$\pm$1.479 &43.277$\pm$1.601 &37.136$\pm$1.263 \\
SWP 43588 &8631.765 &41.661$\pm$1.666 &42.163$\pm$1.560 &35.594$\pm$1.210 \\
SWP 43636 &8635.688 &45.680$\pm$1.827 &49.405$\pm$1.828 &39.628$\pm$1.347 \\
SWP 43676 &8639.862 &53.819$\pm$2.153 &52.490$\pm$1.942 &45.264$\pm$1.539 \\
SWP 43716 &8643.948 &53.593$\pm$2.144 &56.866$\pm$2.104 &47.434$\pm$1.613 \\
SWP 43871 &8649.367 &56.758$\pm$2.270 &55.327$\pm$2.047 &48.963$\pm$1.665 \\
SWP 43872 &8649.454 &55.680$\pm$2.227 &52.634$\pm$1.947 &46.580$\pm$1.584 \\
SWP 43894 &8651.756 &49.667$\pm$1.987 &48.990$\pm$1.813 &43.264$\pm$1.471 \\
SWP 43895 &8651.855 &48.352$\pm$1.934 &52.337$\pm$1.936 &41.435$\pm$1.409 \\
SWP 43921 &8656.130 &61.961$\pm$2.478 &62.436$\pm$2.310 &53.664$\pm$1.825 \\
SWP 43945 &8660.040 &64.188$\pm$2.568 &67.311$\pm$2.491 &51.531$\pm$1.752 \\
SWP 43946 &8660.121 &63.656$\pm$2.546 &63.607$\pm$2.353 &52.300$\pm$1.778 \\
SWP 43962 &8664.048 &43.983$\pm$1.759 &43.716$\pm$1.617 &40.467$\pm$1.376 \\
SWP 43995 &8668.022 &32.648$\pm$1.306 &30.813$\pm$1.140 &31.508$\pm$1.071 \\
SWP 43996 &8668.115 &31.491$\pm$1.260 &32.760$\pm$1.212 &32.328$\pm$1.099 \\
SWP 44020 &8672.106 &40.887$\pm$1.635 &44.671$\pm$1.653 &38.755$\pm$1.318 \\
SWP 44048 &8676.272 &44.897$\pm$1.796 &49.399$\pm$1.828 &44.148$\pm$1.501 \\
SWP 44072 &8680.310 &50.035$\pm$2.001 &49.829$\pm$1.844 &42.224$\pm$1.436 \\
SWP 44099 &8684.199 &49.072$\pm$1.963 &38.191$\pm$1.413 &33.057$\pm$1.124 \\
SWP 44100 &8684.275 &36.331$\pm$1.453 &39.336$\pm$1.455 &32.631$\pm$1.109 \\
SWP 44126 &8688.228 &33.912$\pm$1.356 &27.773$\pm$1.028 &30.118$\pm$1.024 \\
SWP 44149 &8692.216 &28.542$\pm$1.142 &30.410$\pm$1.125 &28.579$\pm$0.972 \\
SWP 44176 &8695.910 &27.667$\pm$1.107 &29.197$\pm$1.080 &26.109$\pm$0.888 \\
SWP 44189 &8699.713 &28.465$\pm$1.139 &27.397$\pm$1.014 &29.131$\pm$0.990 \\
SWP 44208 &8703.723 &27.708$\pm$1.108 &24.374$\pm$0.902 &26.402$\pm$0.898 \\
SWP 44237 &8707.871 &25.508$\pm$1.020 &33.736$\pm$1.248 &31.970$\pm$1.087 \\
SWP 44267 &8711.731 &32.335$\pm$1.293 &45.001$\pm$1.665 &39.240$\pm$1.334 \\
SWP 44307 &8715.612 &44.933$\pm$1.797 &49.245$\pm$1.822 &45.781$\pm$1.557 \\
SWP 44349 &8719.932 &46.150$\pm$1.846 &55.077$\pm$2.038 &47.230$\pm$1.606 \\
SWP 44350 &8720.020 &56.100$\pm$2.244 &54.303$\pm$2.009 &46.392$\pm$1.577 \\
SWP 44381 &8724.004 &50.768$\pm$2.031 &44.383$\pm$1.642 &42.641$\pm$1.450 \\
SWP 44408 &8727.967 &49.801$\pm$1.992 &44.960$\pm$1.664 &41.495$\pm$1.411 \\
SWP 44409 &8728.068 &42.096$\pm$1.684 &43.414$\pm$1.606 &38.242$\pm$1.300 \\
SWP 44410 &8728.168 &45.123$\pm$1.805 &43.196$\pm$1.598 &40.839$\pm$1.389 \\
SWP 44434 &8731.946 &43.086$\pm$1.723 &49.410$\pm$1.828 &46.087$\pm$1.567 \\
SWP 44435 &8732.205 &49.680$\pm$1.987 &48.115$\pm$1.780 &43.598$\pm$1.482 \\
SWP 44461 &8735.946 &48.231$\pm$1.929 &39.557$\pm$1.464 &38.632$\pm$1.313 \\
SWP 44486 &8739.990 &41.475$\pm$1.659 &35.484$\pm$1.313 &35.752$\pm$1.216 \\
SWP 44492 &8744.129 &37.318$\pm$1.493 &47.640$\pm$1.763 &40.569$\pm$1.379 \\
SWP 44581 &8747.874 &44.276$\pm$1.771 &45.235$\pm$1.674 &41.850$\pm$1.423 \\
SWP 44627 &8751.845 &45.411$\pm$1.816 &55.741$\pm$2.062 &47.994$\pm$1.632 \\
SWP 44628 &8751.950 &51.002$\pm$2.040 &53.867$\pm$1.993 &48.992$\pm$1.666 \\
SWP 44629 &8752.056 &53.204$\pm$2.128 &54.219$\pm$2.006 &47.375$\pm$1.611 \\
SWP 44659 &8755.872 &52.803$\pm$2.112 &46.318$\pm$1.714 &45.237$\pm$1.538 \\
SWP 44660 &8755.949 &47.175$\pm$1.887 &44.828$\pm$1.659 &40.003$\pm$1.360 \\
SWP 44682 &8759.702 &47.448$\pm$1.898 &51.934$\pm$1.922 &48.939$\pm$1.664 \\
SWP 44731 &8763.559 &52.863$\pm$2.115 &44.915$\pm$1.662 &42.596$\pm$1.448 \\
SWP 44760 &8767.535 &47.646$\pm$1.906 &45.992$\pm$1.702 &41.306$\pm$1.404 \\
SWP 44803 &8771.689 &47.353$\pm$1.894 &47.380$\pm$1.753 &41.833$\pm$1.422 \\
SWP 44804 &8771.772 &52.010$\pm$2.080 &46.543$\pm$1.722 &42.305$\pm$1.438 \\
SWP 44830 &8775.633 &46.666$\pm$1.867 &44.004$\pm$1.628 &38.184$\pm$1.298 \\
SWP 44873 &8779.607 &54.715$\pm$2.189 &54.094$\pm$2.001 &53.352$\pm$1.814 \\
SWP 44907 &8783.948 &56.424$\pm$2.257 &54.764$\pm$2.026 &49.191$\pm$1.672 \\
SWP 44918 &8785.949 &57.362$\pm$2.294 &55.285$\pm$2.046 &50.312$\pm$1.711 \\
SWP 44921 &8787.786 &57.131$\pm$2.285 &57.471$\pm$2.126 &49.797$\pm$1.693 \\
SWP 44922 &8787.864 &56.597$\pm$2.264 &59.267$\pm$2.193 &46.567$\pm$1.583 \\
SWP 44935 &8790.113 &54.404$\pm$2.176 &60.190$\pm$2.227 &50.947$\pm$1.732 \\
SWP 44949 &8791.769 &60.256$\pm$2.410 &55.655$\pm$2.059 &49.797$\pm$1.693 \\
SWP 44950 &8791.857 &70.247$\pm$2.810 &55.465$\pm$2.052 &55.578$\pm$1.890 \\
SWP 44964 &8793.963 &57.656$\pm$2.306 &56.226$\pm$2.080 &48.742$\pm$1.657 \\
SWP 44974 &8795.768 &56.318$\pm$2.253 &53.796$\pm$1.990 &48.098$\pm$1.635 \\
SWP 44992 &8797.448 &54.836$\pm$2.193 &53.977$\pm$1.997 &46.297$\pm$1.574 \\
SWP 44993 &8797.538 &52.793$\pm$2.112 &54.654$\pm$2.022 &46.258$\pm$1.573 \\
SWP 45010 &8799.460 &54.291$\pm$2.172 &47.346$\pm$1.752 &50.123$\pm$1.704 \\
SWP 45024 &8801.764 &55.805$\pm$2.232 &50.718$\pm$1.877 &45.454$\pm$1.545 \\
SWP 45025 &8801.887 &49.579$\pm$1.983 &54.021$\pm$1.999 &43.688$\pm$1.485 \\
SWP 45026 &8801.996 &44.498$\pm$1.780 &52.914$\pm$1.958 &44.756$\pm$1.522 \\
SWP 45038 &8803.458 &52.139$\pm$2.086 &58.489$\pm$2.164 &49.668$\pm$1.689 \\
SWP 45052 &8805.543 &50.742$\pm$2.030 &61.563$\pm$2.278 &52.663$\pm$1.791 \\
SWP 45063 &8807.520 &51.641$\pm$2.066 &59.254$\pm$2.192 &50.333$\pm$1.711 \\
SWP 45064 &8807.603 &51.587$\pm$2.063 &62.612$\pm$2.317 &54.971$\pm$1.869 \\
SWP 45081 &8809.509 &59.759$\pm$2.390 &57.843$\pm$2.140 &50.158$\pm$1.705 \\
SWP 45082 &8809.601 &60.228$\pm$2.409 &53.138$\pm$1.966 &51.993$\pm$1.768 \\
SWP 45096 &8811.493 &63.860$\pm$2.554 &55.395$\pm$2.050 &49.903$\pm$1.697 \\
SWP 45097 &8811.595 &56.369$\pm$2.255 &52.657$\pm$1.948 &50.072$\pm$1.702 \\
SWP 45106 &8813.384 &51.571$\pm$2.063 &54.998$\pm$2.035 &47.810$\pm$1.626 \\
SWP 45118 &8816.028 &57.179$\pm$2.287 &51.851$\pm$1.918 &45.563$\pm$1.549 \\
SWP 45133 &8818.024 &57.305$\pm$2.292 &42.796$\pm$1.583 &39.882$\pm$1.356 \\
SWP 45150 &8819.700 &54.163$\pm$2.167 &37.296$\pm$1.380 &\nodata \\
SWP 45151 &8819.800 &52.810$\pm$2.112 &38.618$\pm$1.429 &38.881$\pm$1.322 \\
SWP 45152 &8819.904 &41.373$\pm$1.655 &37.208$\pm$1.377 &38.237$\pm$1.300 \\
SWP 45167 &8821.689 &39.568$\pm$1.583 &45.306$\pm$1.676 &44.545$\pm$1.515 \\
SWP 45168 &8821.791 &48.325$\pm$1.933 &47.669$\pm$1.764 &46.635$\pm$1.586 \\
SWP 45169 &8821.892 &51.088$\pm$2.044 &53.395$\pm$1.976 &43.804$\pm$1.489 \\
SWP 45194 &8824.353 &50.394$\pm$2.016 &63.151$\pm$2.337 &52.316$\pm$1.779 \\
SWP 45195 &8824.440 &61.519$\pm$2.461 &60.085$\pm$2.223 &51.566$\pm$1.753 \\
SWP 45206 &8825.701 &60.748$\pm$2.430 &63.864$\pm$2.363 &\nodata \\
SWP 45207 &8825.798 &66.760$\pm$2.670 &62.632$\pm$2.317 &53.268$\pm$1.811 \\
SWP 45219 &8827.904 &67.384$\pm$2.695 &67.817$\pm$2.509 &59.430$\pm$2.021 \\
SWP 45227 &8829.302 &69.254$\pm$2.770 &59.651$\pm$2.207 &56.059$\pm$1.906 \\
SWP 45237 &8831.317 &70.910$\pm$2.836 &67.452$\pm$2.496 &58.335$\pm$1.983 \\
SWP 45246 &8833.326 &72.855$\pm$2.914 &76.033$\pm$2.813 &59.754$\pm$2.032 \\
\enddata
\tablenotetext{a}{Continuum fluxes are given in units of 10$^{-15}$ ergs cm$^{-2}$ s$^{-1}$ \AA$^{-1}$.}
\end{deluxetable}

\clearpage

\begin{deluxetable}{lcccccc}
\tabletypesize{\footnotesize}
\tablecaption{UV Emission Line Flux Data\tablenotemark{a} \label{tab3}}
\tablewidth{0pt}
\tablehead{
\colhead{Image} & \colhead{Julian Date} &
\colhead{He II $\lambda$1640 +} & \colhead{Si IV $\lambda$1400 +} & \colhead{} & 
\colhead{} & \colhead{Si III] $\lambda$1892 +} \\
\colhead{Name} & \colhead{(2,440,000+)} &
\colhead{O III] $\lambda$1663} & \colhead{O IV] $\lambda$1402} & \colhead{\lya} & 
\colhead{C IV $\lambda$1549} & \colhead{C III] $\lambda$1909} 
}
\startdata
SWP 43438 &8611.948 &17.051$\pm$1.705 &12.769$\pm$1.430 &41.751$\pm$1.837 &76.614$\pm$2.988 & 9.707$\pm$1.175 \\
SWP 43439 &8612.031 &14.790$\pm$1.479 &10.856$\pm$1.216 &41.387$\pm$1.821 &74.662$\pm$2.912 & 9.180$\pm$1.111 \\
SWP 43472 &8615.949 &15.837$\pm$1.584 &10.362$\pm$1.161 &38.720$\pm$1.704 &69.633$\pm$2.716 & 8.968$\pm$1.085 \\
SWP 43473 &8616.029 &15.365$\pm$1.536 &11.732$\pm$1.314 &39.614$\pm$1.743 &71.235$\pm$2.778 &13.764$\pm$1.665 \\
SWP 43485 &8618.118 &13.188$\pm$1.319 & 8.559$\pm$0.959 &37.369$\pm$1.644 &73.775$\pm$2.877 &10.270$\pm$1.243 \\
SWP 43539 &8624.202 &14.384$\pm$1.438 &10.623$\pm$1.190 &39.431$\pm$1.735 &66.575$\pm$2.596 &10.175$\pm$1.231 \\
SWP 43540 &8624.294 &17.992$\pm$1.799 & 9.900$\pm$1.109 &36.819$\pm$1.620 &69.660$\pm$2.717 &13.212$\pm$1.599 \\
SWP 43541 &8624.385 &18.288$\pm$1.829 &12.007$\pm$1.345 &39.643$\pm$1.744 &74.851$\pm$2.919 &10.466$\pm$1.266 \\
SWP 43557 &8628.595 &18.169$\pm$1.817 & 9.672$\pm$1.083 &36.259$\pm$1.595 &65.172$\pm$2.542 &11.489$\pm$1.390 \\
SWP 43587 &8631.680 &12.129$\pm$1.213 & 9.326$\pm$1.045 &37.353$\pm$1.644 &68.481$\pm$2.671 &11.690$\pm$1.414 \\
SWP 43588 &8631.765 &11.233$\pm$1.123 & 9.692$\pm$1.086 &35.715$\pm$1.571 &71.312$\pm$2.781 &13.907$\pm$1.683 \\
SWP 43636 &8635.688 &13.382$\pm$1.338 & 9.124$\pm$1.022 &33.320$\pm$1.466 &64.623$\pm$2.520 & 9.733$\pm$1.178 \\
SWP 43676 &8639.862 &17.282$\pm$1.728 &10.188$\pm$1.141 &36.059$\pm$1.587 &68.877$\pm$2.686 &10.664$\pm$1.290 \\
SWP 43716 &8643.948 &18.714$\pm$1.871 &11.037$\pm$1.236 &40.307$\pm$1.774 &73.296$\pm$2.859 &10.240$\pm$1.239 \\
SWP 43871 &8649.367 &13.185$\pm$1.319 &11.574$\pm$1.296 &42.142$\pm$1.854 &72.814$\pm$2.840 &10.015$\pm$1.212 \\
SWP 43872 &8649.454 &15.884$\pm$1.588 &12.645$\pm$1.416 &45.933$\pm$2.021 &76.959$\pm$3.001 &10.892$\pm$1.318 \\
SWP 43894 &8651.756 &12.264$\pm$1.226 &10.932$\pm$1.224 &38.313$\pm$1.686 &67.904$\pm$2.648 &12.443$\pm$1.506 \\
SWP 43895 &8651.855 &14.863$\pm$1.486 & 9.827$\pm$1.101 &38.525$\pm$1.695 &67.918$\pm$2.649 &10.770$\pm$1.303 \\
SWP 43921 &8656.130 &15.008$\pm$1.501 & 9.425$\pm$1.056 &40.596$\pm$1.786 &67.940$\pm$2.650 & 8.425$\pm$1.019 \\
SWP 43945 &8660.040 &14.856$\pm$1.486 &11.559$\pm$1.295 &39.347$\pm$1.731 &70.191$\pm$2.737 &10.595$\pm$1.282 \\
SWP 43946 &8660.121 &13.709$\pm$1.371 &11.282$\pm$1.264 &42.857$\pm$1.886 &68.775$\pm$2.682 & 8.694$\pm$1.052 \\
SWP 43962 &8664.048 &13.022$\pm$1.302 &11.275$\pm$1.263 &40.307$\pm$1.774 &71.699$\pm$2.796 &10.706$\pm$1.295 \\
SWP 43995 &8668.022 &\nodata          & 7.140$\pm$0.800 &35.489$\pm$1.562 &66.840$\pm$2.607 &11.294$\pm$1.367 \\
SWP 43996 &8668.115 &\nodata          & 9.901$\pm$1.109 &35.812$\pm$1.576 &69.053$\pm$2.693 &10.296$\pm$1.246 \\
SWP 44020 &8672.106 &11.764$\pm$1.176 &10.083$\pm$1.129 &36.446$\pm$1.604 &64.753$\pm$2.525 & 9.706$\pm$1.174 \\
SWP 44048 &8676.272 &14.299$\pm$1.430 &10.509$\pm$1.177 &35.112$\pm$1.545 &60.287$\pm$2.351 &10.137$\pm$1.227 \\
SWP 44072 &8680.310 &11.451$\pm$1.145 & 9.171$\pm$1.027 &33.214$\pm$1.461 &65.161$\pm$2.541 & 9.986$\pm$1.208 \\
SWP 44099 &8684.199 &13.925$\pm$1.393 &10.682$\pm$1.196 &33.220$\pm$1.462 &67.461$\pm$2.631 &11.558$\pm$1.399 \\
SWP 44100 &8684.275 &11.177$\pm$1.118 &12.329$\pm$1.381 &31.722$\pm$1.396 &64.203$\pm$2.504 &14.211$\pm$1.720 \\
SWP 44126 &8688.228 & 8.477$\pm$0.848 & 7.211$\pm$0.808 &29.611$\pm$1.303 &57.594$\pm$2.246 & 9.462$\pm$1.145 \\
SWP 44149 &8692.216 & 8.912$\pm$0.891 & 7.930$\pm$0.888 &29.519$\pm$1.299 &63.911$\pm$2.493 & 9.846$\pm$1.191 \\
SWP 44176 &8695.910 &10.943$\pm$1.094 & 8.069$\pm$0.904 &26.618$\pm$1.171 &58.785$\pm$2.293 &10.981$\pm$1.329 \\
SWP 44189 &8699.713 & 8.945$\pm$0.894 & 8.916$\pm$0.999 &28.493$\pm$1.254 &57.753$\pm$2.252 & 7.723$\pm$0.934 \\
SWP 44208 &8703.723 & 7.663$\pm$0.766 & 8.471$\pm$0.949 &24.982$\pm$1.099 &53.504$\pm$2.087 & 8.793$\pm$1.064 \\
SWP 44237 &8707.871 & 9.909$\pm$0.991 & 8.974$\pm$1.005 &27.480$\pm$1.209 &56.975$\pm$2.222 & 7.860$\pm$0.951 \\
SWP 44267 &8711.731 &13.391$\pm$1.339 & 9.154$\pm$1.025 &31.722$\pm$1.396 &60.820$\pm$2.372 & 8.547$\pm$1.034 \\
SWP 44307 &8715.612 &11.593$\pm$1.159 &12.355$\pm$1.384 &33.015$\pm$1.453 &65.884$\pm$2.569 & 9.412$\pm$1.139 \\
SWP 44349 &8719.932 &11.976$\pm$1.198 & 9.588$\pm$1.074 &33.504$\pm$1.474 &61.410$\pm$2.395 & 9.790$\pm$1.185 \\
SWP 44350 &8720.020 &15.663$\pm$1.566 &13.091$\pm$1.466 &36.571$\pm$1.609 &67.330$\pm$2.626 &10.934$\pm$1.323 \\
SWP 44381 &8724.004 &12.488$\pm$1.249 & 8.475$\pm$0.949 &38.718$\pm$1.704 &64.045$\pm$2.498 & 9.589$\pm$1.160 \\
SWP 44408 &8727.967 &11.592$\pm$1.159 &11.942$\pm$1.338 &39.422$\pm$1.735 &68.315$\pm$2.664 & 7.924$\pm$0.959 \\
SWP 44409 &8728.068 &\nodata          &11.346$\pm$1.271 &37.210$\pm$1.637 &67.746$\pm$2.642 &11.943$\pm$1.445 \\
SWP 44410 &8728.168 &15.773$\pm$1.577 &11.716$\pm$1.312 &37.375$\pm$1.645 &72.573$\pm$2.830 &11.679$\pm$1.413 \\
SWP 44434 &8731.946 &16.459$\pm$1.646 &11.614$\pm$1.301 &41.213$\pm$1.813 &72.594$\pm$2.831 & 8.378$\pm$1.014 \\
SWP 44435 &8732.205 &\nodata          &13.598$\pm$1.523 &40.652$\pm$1.789 &68.736$\pm$2.681 & 9.980$\pm$1.208 \\
SWP 44461 &8735.946 &14.037$\pm$1.404 &10.790$\pm$1.208 &37.320$\pm$1.642 &69.492$\pm$2.710 &11.347$\pm$1.373 \\
SWP 44486 &8739.990 &10.913$\pm$1.091 &11.056$\pm$1.238 &33.716$\pm$1.484 &63.079$\pm$2.460 &10.973$\pm$1.328 \\
SWP 44492 &8744.129 &10.972$\pm$1.097 & 8.646$\pm$0.968 &31.276$\pm$1.376 &63.631$\pm$2.482 & 9.638$\pm$1.166 \\
SWP 44581 &8747.874 &13.851$\pm$1.385 &10.644$\pm$1.192 &37.659$\pm$1.657 &71.354$\pm$2.783 & 9.569$\pm$1.158 \\
SWP 44627 &8751.845 &15.350$\pm$1.535 &10.311$\pm$1.155 &34.631$\pm$1.524 &72.491$\pm$2.827 & 8.062$\pm$0.976 \\
SWP 44628 &8751.950 &14.578$\pm$1.458 &10.779$\pm$1.207 &39.274$\pm$1.728 &71.496$\pm$2.788 & 9.285$\pm$1.123 \\
SWP 44629 &8752.056 &12.920$\pm$1.292 & 8.708$\pm$0.975 &36.988$\pm$1.627 &67.524$\pm$2.633 & 9.705$\pm$1.174 \\
SWP 44659 &8755.872 &13.868$\pm$1.387 &10.749$\pm$1.204 &37.794$\pm$1.663 &68.993$\pm$2.691 & 9.236$\pm$1.118 \\
SWP 44660 &8755.949 &\nodata          & 9.701$\pm$1.087 &37.884$\pm$1.667 &71.405$\pm$2.785 &13.026$\pm$1.576 \\
SWP 44682 &8759.702 &13.730$\pm$1.373 &11.811$\pm$1.323 &37.249$\pm$1.639 &68.033$\pm$2.653 & 7.400$\pm$0.895 \\
SWP 44731 &8763.559 &15.880$\pm$1.588 &10.974$\pm$1.229 &36.521$\pm$1.607 &65.598$\pm$2.558 & 9.866$\pm$1.194 \\
SWP 44760 &8767.535 &10.554$\pm$1.055 & 9.223$\pm$1.033 &31.337$\pm$1.379 &54.612$\pm$2.130 &10.060$\pm$1.217 \\
SWP 44803 &8771.689 &12.725$\pm$1.273 &11.678$\pm$1.308 &30.261$\pm$1.331 &60.436$\pm$2.357 & 9.198$\pm$1.113 \\
SWP 44804 &8771.772 &12.169$\pm$1.217 & 8.260$\pm$0.925 &30.620$\pm$1.347 &63.469$\pm$2.475 &12.367$\pm$1.496 \\
SWP 44830 &8775.633 &15.957$\pm$1.596 &10.018$\pm$1.122 &34.556$\pm$1.520 &61.264$\pm$2.389 &11.703$\pm$1.416 \\
SWP 44873 &8779.607 &12.230$\pm$1.223 &13.309$\pm$1.491 &33.902$\pm$1.492 &62.857$\pm$2.451 & 7.113$\pm$0.861 \\
SWP 44907 &8783.948 &15.930$\pm$1.593 & 9.829$\pm$1.101 &35.165$\pm$1.547 &65.009$\pm$2.535 & 8.760$\pm$1.060 \\
SWP 44918 &8785.949 &10.634$\pm$1.063 &11.991$\pm$1.343 &36.745$\pm$1.617 &62.983$\pm$2.456 & 6.630$\pm$0.802 \\
SWP 44921 &8787.786 &12.702$\pm$1.270 &10.055$\pm$1.126 &34.865$\pm$1.534 &64.779$\pm$2.526 &10.548$\pm$1.276 \\
SWP 44922 &8787.864 &11.840$\pm$1.184 & 8.687$\pm$0.973 &34.194$\pm$1.505 &66.079$\pm$2.577 &11.881$\pm$1.438 \\
SWP 44935 &8790.113 &14.784$\pm$1.478 & 9.681$\pm$1.084 &33.982$\pm$1.495 &67.992$\pm$2.652 &11.232$\pm$1.359 \\
SWP 44949 &8791.769 &15.495$\pm$1.549 &11.760$\pm$1.317 &37.262$\pm$1.640 &70.532$\pm$2.751 & 9.822$\pm$1.188 \\
SWP 44950 &8791.857 &15.004$\pm$1.500 &10.425$\pm$1.168 &36.716$\pm$1.616 &67.412$\pm$2.629 &14.068$\pm$1.702 \\
SWP 44964 &8793.963 &14.658$\pm$1.466 & 9.572$\pm$1.072 &37.016$\pm$1.629 &67.780$\pm$2.643 &10.045$\pm$1.215 \\
SWP 44974 &8795.768 &14.577$\pm$1.458 &10.878$\pm$1.218 &37.405$\pm$1.646 &69.919$\pm$2.727 & 8.464$\pm$1.024 \\
SWP 44992 &8797.448 &16.789$\pm$1.679 &11.013$\pm$1.233 &37.606$\pm$1.655 &71.690$\pm$2.796 &11.005$\pm$1.332 \\
SWP 44993 &8797.538 &15.144$\pm$1.514 &10.938$\pm$1.225 &37.047$\pm$1.630 &71.722$\pm$2.797 &10.318$\pm$1.248 \\
SWP 45010 &8799.460 &15.066$\pm$1.507 &10.861$\pm$1.216 &38.602$\pm$1.698 &61.098$\pm$2.383 & 9.052$\pm$1.095 \\
SWP 45024 &8801.764 &11.581$\pm$1.158 &11.610$\pm$1.300 &\nodata          &63.585$\pm$2.480 &11.248$\pm$1.361 \\
SWP 45025 &8801.887 &13.310$\pm$1.331 & 9.372$\pm$1.050 &34.032$\pm$1.497 &64.283$\pm$2.507 &10.079$\pm$1.220 \\
SWP 45026 &8801.996 &11.679$\pm$1.168 & 9.878$\pm$1.106 &35.559$\pm$1.565 &64.598$\pm$2.519 &11.732$\pm$1.420 \\
SWP 45038 &8803.458 &11.433$\pm$1.143 &11.278$\pm$1.263 &36.813$\pm$1.620 &66.819$\pm$2.606 & 8.374$\pm$1.013 \\
SWP 45052 &8805.543 &16.844$\pm$1.684 &12.541$\pm$1.405 &41.348$\pm$1.819 &73.452$\pm$2.865 &10.795$\pm$1.306 \\
SWP 45063 &8807.520 &16.108$\pm$1.611 &13.618$\pm$1.525 &40.783$\pm$1.794 &72.034$\pm$2.809 &11.587$\pm$1.402 \\
SWP 45064 &8807.603 &15.292$\pm$1.529 &12.634$\pm$1.415 &36.449$\pm$1.604 &72.370$\pm$2.822 &12.714$\pm$1.538 \\
SWP 45081 &8809.509 &14.375$\pm$1.438 &12.532$\pm$1.404 &35.314$\pm$1.554 &70.550$\pm$2.751 &11.009$\pm$1.332 \\
SWP 45082 &8809.601 &14.549$\pm$1.455 &13.052$\pm$1.462 &38.033$\pm$1.673 &74.732$\pm$2.915 & 9.878$\pm$1.195 \\
SWP 45096 &8811.493 &15.791$\pm$1.579 &13.392$\pm$1.500 &35.964$\pm$1.582 &69.099$\pm$2.695 &11.158$\pm$1.350 \\
SWP 45097 &8811.595 &15.899$\pm$1.590 &13.384$\pm$1.499 &38.451$\pm$1.692 &72.306$\pm$2.820 &10.461$\pm$1.266 \\
SWP 45106 &8813.384 &14.546$\pm$1.455 &10.726$\pm$1.201 &36.440$\pm$1.603 &64.005$\pm$2.496 &12.925$\pm$1.564 \\
SWP 45118 &8816.028 &13.700$\pm$1.370 & 8.430$\pm$0.944 &34.493$\pm$1.518 &67.487$\pm$2.632 &11.906$\pm$1.441 \\
SWP 45133 &8818.024 &10.933$\pm$1.093 &11.187$\pm$1.253 &32.175$\pm$1.416 &66.611$\pm$2.598 &11.225$\pm$1.358 \\
SWP 45150 &8819.700 & 8.954$\pm$0.895 & 8.490$\pm$0.951 &27.355$\pm$1.204 &55.596$\pm$2.168 &\nodata \\         
SWP 45151 &8819.800 & 7.794$\pm$0.779 & 6.574$\pm$0.736 &31.274$\pm$1.376 &52.838$\pm$2.061 &12.267$\pm$1.484 \\
SWP 45152 &8819.904 & 7.073$\pm$0.707 & 8.813$\pm$0.987 &30.107$\pm$1.325 &59.321$\pm$2.314 & 9.929$\pm$1.201 \\
SWP 45167 &8821.689 &11.037$\pm$1.104 & 8.706$\pm$0.975 &30.555$\pm$1.344 &62.589$\pm$2.441 &10.167$\pm$1.230 \\
SWP 45168 &8821.791 &12.697$\pm$1.270 & 7.656$\pm$0.857 &29.693$\pm$1.306 &57.353$\pm$2.237 & 9.218$\pm$1.115 \\
SWP 45169 &8821.892 &11.109$\pm$1.111 & 8.925$\pm$1.000 &29.510$\pm$1.298 &56.358$\pm$2.198 &11.512$\pm$1.393 \\
SWP 45194 &8824.353 &14.707$\pm$1.471 & 9.936$\pm$1.113 &31.339$\pm$1.379 &59.284$\pm$2.312 & 8.923$\pm$1.080 \\
SWP 45195 &8824.440 &13.914$\pm$1.391 &\nodata          &29.242$\pm$1.287 &61.297$\pm$2.391 & 9.275$\pm$1.122 \\
SWP 45206 &8825.701 &14.222$\pm$1.422 &10.173$\pm$1.139 &33.615$\pm$1.479 &64.641$\pm$2.521 &\nodata \\         
SWP 45207 &8825.798 &16.205$\pm$1.620 & 9.322$\pm$1.044 &30.904$\pm$1.360 &60.589$\pm$2.363 &10.583$\pm$1.281 \\
SWP 45219 &8827.904 &14.897$\pm$1.490 &12.322$\pm$1.380 &35.942$\pm$1.581 &63.532$\pm$2.478 & 8.180$\pm$0.990 \\
SWP 45227 &8829.302 &14.370$\pm$1.437 &14.745$\pm$1.651 &36.139$\pm$1.590 &72.627$\pm$2.832 & 9.507$\pm$1.150 \\
SWP 45237 &8831.317 &14.427$\pm$1.443 &13.534$\pm$1.516 &35.966$\pm$1.583 &66.867$\pm$2.608 & 9.803$\pm$1.186 \\
SWP 45246 &8833.326 &18.615$\pm$1.862 &11.108$\pm$1.244 &36.969$\pm$1.627 &69.398$\pm$2.707 &11.312$\pm$1.369 \\
\enddata
\tablenotetext{a}{Emission line fluxes are in units of 10$^{-13}$ ergs cm$^{-2}$ s$^{-1}$.}
\end{deluxetable}

\clearpage

\begin{deluxetable}{lcc}
\tablecaption{UV Velocity Data\tablenotemark{a} \label{tab4}}
\tablewidth{0pt}
\tablehead{
\colhead{Emission Line} & \colhead{V$^{rest}_{FWHM}$(rms)} & \colhead{V$^{rest}_{FWHM}$(mean)}
}
\startdata
\heo &6.34$\pm$0.90 &4.74$\pm$0.69 \\
\sio &5.73$\pm$2.71 &4.81$\pm$0.50 \\
\lya &\nodata &\nodata \\
\car &3.55$\pm$0.59 &3.03$\pm$0.07 \\
\sic &2.61$\pm$0.16 &2.82$\pm$0.31 \\
\enddata
\tablenotetext{a}{Velocity data are in units of 10$^{3}$ km s$^{-1}$.}
\end{deluxetable}

\clearpage

\begin{deluxetable}{cccc}
\tablecaption{Optical Flux Data \label{tab5}}
\tablewidth{0pt}
\tablehead{
\colhead{Image} & \colhead{Julian Date} & \colhead{} & \colhead{} \\
\colhead{Name} & \colhead{(2,440,000+)} & \colhead{F($\lambda5150$)\tablenotemark{a}} & 
\colhead{\hb\tablenotemark{b}}
}
\startdata
n38607a &8607.830 &11.406$\pm$0.605 & 9.829$\pm$0.403 \\
n38623a &8623.830 &11.707$\pm$0.620 &10.100$\pm$0.414 \\
n38627a &8627.830 &11.094$\pm$0.588 &10.272$\pm$0.421 \\
n38631a &8631.840 &10.865$\pm$0.576 &10.494$\pm$0.430 \\
n38635a &8635.830 &10.781$\pm$0.571 &10.183$\pm$0.418 \\
n38639a &8639.840 &11.303$\pm$0.599 &10.327$\pm$0.423 \\
n38643a &8643.720 &12.788$\pm$0.678 &10.423$\pm$0.427 \\
n38647a &8647.760 &12.668$\pm$0.671 &10.459$\pm$0.429 \\
n38651a &8651.670 &11.190$\pm$0.593 &10.603$\pm$0.435 \\
n38656a &8656.770 &12.163$\pm$0.645 &10.669$\pm$0.437 \\
n38660a &8660.770 &11.629$\pm$0.616 &10.109$\pm$0.414 \\
n38664a &8664.770 &11.005$\pm$0.583 &10.894$\pm$0.447 \\
n38668a &8668.810 &10.204$\pm$0.541 &10.969$\pm$0.450 \\
n38676a &8676.710 &10.817$\pm$0.573 & 9.760$\pm$0.400 \\
n38677a &8677.800 &10.786$\pm$0.572 & 9.483$\pm$0.389 \\
n38678a &8678.680 &10.703$\pm$0.567 & 9.513$\pm$0.390 \\
n38704a &8704.580 & 8.763$\pm$0.464 & 8.469$\pm$0.347 \\ 
n38712a &8712.590 &10.350$\pm$0.549 & 8.410$\pm$0.345 \\
n38716a &8716.600 &10.055$\pm$0.533 & 8.840$\pm$0.362 \\
n38720a &8720.590 &11.842$\pm$0.628 & 9.395$\pm$0.385 \\
n38724a &8724.560 &11.257$\pm$0.597 & 9.640$\pm$0.395 \\
n38732a &8732.560 &11.212$\pm$0.594 & 9.357$\pm$0.384 \\
n38736a &8736.550 &10.149$\pm$0.538 & 9.410$\pm$0.386 \\
n38744a &8744.610 &10.812$\pm$0.573 &10.537$\pm$0.432 \\
n38752a &8752.570 &11.174$\pm$0.592 & 9.606$\pm$0.394 \\
n38763a &8763.580 &11.880$\pm$0.630 &11.678$\pm$0.479 \\
n38764a &8764.560 &10.754$\pm$0.570 &10.131$\pm$0.415 \\
n38772a &8772.630 &11.799$\pm$0.625 & 9.627$\pm$0.395 \\
n38776a &8776.560 &10.636$\pm$0.564 & 9.272$\pm$0.380 \\
n38794a &8794.570 &11.444$\pm$0.607 &10.803$\pm$0.443 \\
n38804a &8804.470 &12.281$\pm$0.651 &11.780$\pm$0.483 \\
n38808a &8808.480 &12.773$\pm$0.677 &10.531$\pm$0.432 \\
n38822a &8822.480 &12.917$\pm$0.685 &11.060$\pm$0.453 \\
n38824a &8824.470 &11.765$\pm$0.624 &10.108$\pm$0.414 \\
n38826a &8826.470 &12.717$\pm$0.674 & 9.912$\pm$0.406 \\
n38830a &8830.480 &12.108$\pm$0.642 &11.026$\pm$0.452 \\
n38832a &8832.460 &13.281$\pm$0.704 &11.086$\pm$0.455 \\
\enddata
\tablenotetext{a}{Continuum fluxes are in units of 10$^{-15}$ ergs cm$^{-2}$ s$^{-1}$ \AA$^{-1}$.}
\tablenotetext{b}{Emission line fluxes are in units of 10$^{-13}$ ergs cm$^{-2}$ s$^{-1}$.}
\end{deluxetable}

\clearpage

\begin{deluxetable}{lcccccc}
\tabletypesize{\footnotesize}
\tablecaption{Sampling Statistics \label{tab6}}
\tablewidth{0pt}
\tablehead{
\colhead{} & \colhead{} & \multicolumn{2}{c}{Sampling Interval (days)} &
\colhead{} & \colhead{} & \colhead{} \\
 \cline{3-4} \\
\colhead{Subset} & \colhead{Number} & \colhead{Average} & \colhead{Median} &
\colhead{F$_{var}$} & \colhead{R$_{max}$} & \colhead{Reference}
}
\startdata
Previous 1460 \AA\ continuum dataset & 69 & 3.3 & 3.9 & 0.201 & 3.027$\pm$0.380 & 1 \\
New UV dataset; binned by epoch & 69 & 3.3 & 3.9 & 0.203 & 3.119$\pm$0.163 \\
New UV dataset; complete sample & 101 & 2.2 & 2.0 & 0.192 & 2.856$\pm$0.162 \\
New UV dataset; 4-day sampling period & 62 & 2.8 & 3.9 & 0.193 & 2.762$\pm$0.145 \\
New UV dataset; 2-day sampling period & 40 & 1.3 & 1.7 & 0.140 & 2.043$\pm$0.107 \\
Previous 5150 \AA\ continuum dataset & 72 & 3.2 & 2.0 & 0.078 & 1.517$\pm$0.042 & 2 \\
New optical dataset; binned by epoch & 35 & 6.6 & 4.0 & 0.065 & 1.516$\pm$0.114 \\
New optical dataset; complete sample & 37 & 6.2 & 4.0 & 0.064 & 1.516$\pm$0.114 \\
\enddata
\tablerefs{(1) Reichert et al. 1994; (2) Stirpe et al. 1994.}
\tablecomments{Previously published light curves were collected from the AGN Watch website.}
\end{deluxetable}

\clearpage

\begin{deluxetable}{lrrrr}
\tablecaption{Cross-Correlation Results\tablenotemark{a} \label{tab7}}
\tablewidth{0pt}
\tablehead{
\colhead{} & \multicolumn{2}{c}{Previous Results\tablenotemark{b}} & \multicolumn{2}{c}{Current Results} \\ 
\cline{2-3} \cline{4-5} \\ 
\colhead{Line/Band} & \colhead{$\tau_{cent}^{rest}$} & \colhead{$\tau_{peak}^{rest}$} & 
\colhead{$\tau_{cent}^{rest}$} & \colhead{$\tau_{peak}^{rest}$}
}
\startdata
F($\lambda1460$)
&\nodata       &\nodata      &$-$0.1$^{+0.3}_{-0.2}$ &0.0$^{+0.2}_{-0.4}$ \\
F($\lambda1835$)
&0.1$^{+3}_{-3}$ &0$^{+2}_{-2}$ &0.0$^{+0.3}_{-0.3}$ &0.0$^{+0.6}_{-0.5}$ \\
F($\lambda5150$) 
&1.6$^{+2}_{-2}$ &1$^{+2}_{-2}$ &0.4$^{+3.1}_{-1.6}$ &0.7$^{+1.9}_{-1.6}$ \\
He II $\lambda$1640 + O III] $\lambda$1663 
&0.5$^{+4}_{-4}$ &1$^{+2}_{-2}$ &1.3$^{+0.9}_{-0.5}$ &1.4$^{+0.6}_{-1.1}$ \\
Si IV $\lambda$1400 + O IV] $\lambda$1402 
&3.9$^{+4}_{-4}$ &5$^{+2}_{-2}$ &2.1$^{+0.9}_{-1.5}$ &2.3$^{+0.8}_{-2.4}$ \\
Ly$\alpha$
&3.8$^{+3}_{-3}$ &4$^{+2}_{-2}$ &3.6$^{+1.1}_{-0.7}$ &2.2$^{+2.5}_{-0.1}$ \\
C IV $\lambda$1549
&5.4$^{+3}_{-3}$ &5$^{+2}_{-2}$ &3.8$^{+1.0}_{-0.9}$ &4.5$^{+0.4}_{-2.2}$ \\
Si III] $\lambda$1892 + C III] $\lambda$1909\tablenotemark{c}
&15.6$^{+4}_{-4}$ &9$^{+2}_{-2}$ &8.5$^{+1.3}_{-2.6}$ &10.2$^{+0.2}_{-5.3}$ \\
H$\beta$  
&7.1$^{+2}_{-2}$ &8$^{+2}_{-2}$ &10.4$^{+4.1}_{-2.3}$ &9.0$^{+5.1}_{-2.4}$ \\
\enddata
\tablenotetext{a}{All time lag data are in units of days.}
\tablenotetext{b}{The previous results listed here are adapted from the GEX-extracted
UV data of \citet{rei94} and from the optical results of \citet{sti94}, both
of which used the 1460 \AA\ continuum as the driving light curve.}
\tablenotetext{c}{The range of time lags we included in our analysis of \sic\ was limited to 
$\pm$16 days to avoid aliasing.}
\end{deluxetable}

\end{document}